\documentclass[conference]{IEEEtran}
\IEEEoverridecommandlockouts
\usepackage{cite}
\usepackage{amsmath,amssymb,amsfonts}
\usepackage{algorithmic}
\usepackage{graphicx}
\usepackage{textcomp}
\usepackage{xcolor}
\usepackage{marvosym}

\usepackage{bm}
\usepackage{svg} 
\usepackage{booktabs} 
\usepackage{colortbl} 

\def\BibTeX{{\rm B\kern-.05em{\sc i\kern-.025em b}\kern-.08em
    T\kern-.1667em\lower.7ex\hbox{E}\kern-.125emX}}
\begin{document}

\title{Deep Reinforcement Learning for \\ Quantitative Trading}

\author{\IEEEauthorblockN{Maochun Xu, Zixun Lan, Zheng Tao, Jiawei Du, Zongao Ye*}
\IEEEauthorblockA{\textit{Department of Financial and Actuarial Mathematics, School of Mathematics and Physics, Xi’an Jiaotong-Liverpool University} \\
Suzhou 215123, P.R. ChinaEmail:Hong.Yi@xjtlu.edu.cn, Telephone: +86-512-88161729.}
}

\author{\IEEEauthorblockN{1\textsuperscript{st} Maochun Xu}
\IEEEauthorblockA{\textit{dept. Financial and Actuarial Mathematics } \\
\textit{Xi’an Jiaotong-Liverpool University}\\
Suzhou, China \\
Maochun.Xu16@student.xjtlu.edu.cn}
\and
\IEEEauthorblockN{2\textsuperscript{nd} Zixun Lan}
\IEEEauthorblockA{\textit{dept. Applied Mathematics} \\
\textit{Xi’an Jiaotong-Liverpool University}\\
Suzhou, China \\
Zixun.Lan19@student.xjtlu.edu.cn}
\and
\IEEEauthorblockN{3\textsuperscript{rd} Zheng Tao}
\IEEEauthorblockA{\textit{dept. Financial and Actuarial Mathematics } \\
\textit{Xi’an Jiaotong-Liverpool University}\\
Suzhou, China \\
Zheng.Tao22@student.xjtlu.edu.cn}
\and
\IEEEauthorblockN{4\textsuperscript{th} Jiawei Du}
\IEEEauthorblockA{\textit{dept. Financial and Actuarial Mathematics } \\
\textit{Xi’an Jiaotong-Liverpool University}\\
Suzhou, China \\
Jiawei.Du19@student.xjtlu.edu.cn}
\and

\IEEEauthorblockN{}
\IEEEauthorblockA{\textit{\quad\quad\quad\quad\quad\quad\quad\quad\quad\quad\quad\quad\quad} \\
\textit{      \quad         \quad      }\\
\\
    \quad    \quad       \quad    }
\and

\IEEEauthorblockN{5\textsuperscript{th} Zongao Ye$^{\textrm{\Letter}}$}
\IEEEauthorblockA{\textit{dept. Applied Mathematics} \\
\textit{Xi’an Jiaotong-Liverpool University}\\
Suzhou, China \\
Zongao.Ye19@student.xjtlu.edu.cn}

}

\maketitle

\begin{abstract}
Artificial Intelligence (AI) and Machine Learning (ML) are transforming the domain of Quantitative Trading (QT) through the deployment of advanced algorithms capable of sifting through extensive financial datasets to pinpoint lucrative investment openings. AI-driven models, particularly those employing ML techniques such as deep learning and reinforcement learning, have shown great prowess in predicting market trends and executing trades at a speed and accuracy that far surpass human capabilities. Its capacity to automate critical tasks, such as discerning market conditions and executing trading strategies, has been pivotal. However, persistent challenges exist in current QT methods, especially in effectively handling noisy and high-frequency financial data. Striking a balance between exploration and exploitation poses another challenge for AI-driven trading agents. To surmount these hurdles, our proposed solution, QTNet, introduces an adaptive trading model that autonomously formulates QT strategies through an intelligent trading agent. Incorporating deep reinforcement learning (DRL) with imitative learning methodologies, we bolster the proficiency of our model. To tackle the challenges posed by volatile financial datasets, we conceptualize the QT mechanism within the framework of a Partially Observable Markov Decision Process (POMDP). Moreover, by embedding imitative learning, the model can capitalize on traditional trading tactics, nurturing a balanced synergy between discovery and utilization. For a more realistic simulation, our trading agent undergoes training using minute-frequency data sourced from the live financial market. Experimental findings underscore the model's proficiency in extracting robust market features and its adaptability to diverse market conditions.
\end{abstract}


\section{Introduction}
In the realm of financial security investment, quantitative trading (QT) is distinguished by its substantial automation, utilizing computing technology to diminish dependence on personal discretion and mitigate illogical decision-making. As the dominance of quantitative hedge funds grows, there is an increasing focus on integrating machine learning into QT, especially in the context of Fintech. These technologies enable the creation of dynamic trading strategies that can adapt to market changes in real time, manage risks more effectively, and ultimately enhance the profitability and efficiency of trading operations. As the financial markets become increasingly complex, the integration of AI and ML in QT is becoming indispensable for maintaining competitive advantage.

In the fluctuating environment of financial markets, trading behaviors, and economic events are inherently unpredictable, leading to the generation of volatile and non-stationary data. Despite technical analysis being a widely used methodology in Quantitative Trading (QT), its capacity for generalization has been questioned, underscoring the urgency for more resilient features that can be directly mined from raw financial data. As a response, machine learning techniques, especially deep learning models, have been investigated for their potential to predict market trends and improve generalization. Nonetheless, the scope of QT extends well beyond predicting trends; it necessitates the formulation of strategic trading methods. Although reinforcement learning (RL) provides a methodical framework for tackling tasks that involve a series of decisions, achieving an equilibrium between the discovery of novel strategies and the utilization of established ones poses a considerable challenge, especially in the context of the pragmatic constraints encountered in actual trading scenarios.

In response to these difficulties, we introduce Observational and Recurrent Deterministic Policy Gradients (QTNet). We cast QT within the framework of a Partially Observable Markov Decision Process (POMDP) to effectively tackle the representation of unpredictable financial data. Recurrent neural networks, such as Recurrent Deterministic Policy Gradient (RDPG), are employed to handle the POMDP, offering an off-policy deep reinforcement learning (DRL) algorithm.

Balancing exploration and exploitation is addressed through imitative learning techniques. A demonstration buffer, initialized with actions from Dual Thrust, and behavior cloning is introduced to guide the trading agent. By integrating these techniques into the POMDP framework, QTNet benefits from enhanced financial domain knowledge. Real financial data from the futures market is used to test QTNet, demonstrating its ability to learn profitable trading policies and exhibit superior generalization across various futures markets.

\section{Related work}
Research within our field typically falls into two primary classifications. The first, extensively documented by Murphy in 1999, is Quantitative Trading (QT), which is fundamentally dependent on Technical Analysis. This method assumes that all pertinent market details are encoded within the price and volume figures, utilizing a variety of technical indicators—mathematically derived instruments—to signal trading actions. Although these indicators are prevalent, their rigidity often hinders their ability to conform to the market's multifaceted and evolving patterns. Within this realm of indicators, two dominant types exist moving averages, which aim to discern the direction of price trends, and oscillators, designed to gauge the market's momentum. Each type, however, has its inherent constraints as they are based on historical data, which might not always be a reliable predictor of future market behavior due to the dynamic nature of financial markets. This reliance on past data to forecast future trends often fails to accommodate the unpredictable fluctuations that characterize market movements, leading to a demand for more adaptable and nuanced trading tools.

Lately, the integration of machine learning techniques into the realm of securities investment has seen a notable surge, with particular emphasis on Reinforcement Learning (RL) for its aptitude in tackling sequential decision-making tasks. The exploration of RL within the context of Quantitative Trading (QT) is not new; Littman (1996) \cite{littman1996reinforcement, vecerik2017leveraging} was a pioneer in applying Q-learning, a conventional value-based RL algorithm. The landscape of complex problem spaces in QT has acknowledged the limitations of traditional methods. In response, Murphy (1999) \cite{murphy1999technical, lan2021sub, lan2023aednet, lan2022more} and his colleagues, around the turn of the millennium, shifted the focus toward policy-based RL approaches. They put forth the concept of recurrent reinforcement learning (RRL), a method better suited to handle the intricacies and temporal dependencies inherent in financial decision-making processes. This shift underscored the evolving nature of machine learning in finance, seeking algorithms that could not only predict but also learn and adapt strategies over time, harnessing the vast amounts of data to navigate the oft-turbulent waters of the stock market.

Traditional RL faces challenges in selecting appropriate market features. Deep Reinforcement Learning (DRL), combining RL and deep learning, is well-suited for high-dimensional data problems. DRL has shown advancements in tackling complex tasks, including video games, and extends its potential to QT. Jiang, Xu, and Liang (2017)\cite{jiang2017deep, lan2023rcsearcher} leveraged Deep Deterministic Policy Gradient (DDPG) for cryptocurrency portfolios. Efforts have also been made to enhance financial signal representation using fuzzy learning and deep neural networks (DNNs).

Model-free reinforcement learning algorithms, while effective, often struggle with sampling efficiency in extensive state spaces, a situation frequently encountered in QT. To address this, Yu et al. (2019) \cite{yu2019model} and Lan et al. (2023) \cite{zhu2023use} developed a model-based RL approach for daily frequency portfolio management. However, this framework doesn't fully cater to the needs of minute-frequency data, which is more common in QT. To bridge this gap, our study proposes a policy-based RL mechanism that operates in continuous time and is augmented with Recurrent Neural Networks (RNNs). This advanced DRL architecture is tailored to grasp the subtle nuances of minute-by-minute financial data and exhibits flexibility in various financial market conditions.

\section{Problem Definition}
Within this section, we initially provide clear definitions for mathematical symbols, followed by a comprehensive formal introduction to expound upon the intricacies of the quantitative trading problem.

\subsection{Foundations} 

For each discrete moment $t$, we define the OHLC price array as $\bm{p_t}=\left[o_t^p, h_t^p, l_t^p, c_t^p\right]$, where $o_t^p, h_t^p, l_t^p$, and $c_t^p$ correspondingly represent the opening, highest, lowest, and closing prices. The comprehensive price array for a financial instrument is expressed as $\bm{P}=\left[\bm{p_1}, \ldots, \bm{p_t}, \ldots\right]$. For convenience, $\bm{P}_{t-n: t}$ symbolizes the historical price array within a given timeframe, with $n$ denoting the length of the look-back period or window length. The technical indicator vector at time $t$ is denoted as $\bm{I}_t=\left[I_{1 t}, \ldots, I_{j t}\right]$, where $I_{j t}$ is a function of the historical price sequence $\bm{P}_{t-n: t}$ through $I_{j t}=f\left(\bm{P}_{t-n: t} ; \theta_j\right)$, with $\theta_j$ representing the parameters for the technical strategy $j$. The sequence of technical indicators is $\bm{I}=\left[I_1, \ldots, I_t, \ldots\right]$. Similarly, the account profit at time step $t$ is $r_t$, and the sequence of account profits is $\bm{R}=\left[r_1, \ldots, r_t, \ldots\right]$.

\subsection{POMDP}
In this section, we explore the distinctive attributes of Quantitative Trading (QT) and elaborate on the rationale behind framing the entire QT process as a Partially Observable Markov Decision Process (POMDP). The financial market is characterized by security prices influenced by both bullish and bearish investors, macroeconomic and microeconomic activities, unpredictable occurrences, and diverse trading behaviors, contributing to inherent noise. Direct observation of true market states becomes unattainable due to these complexities. For instance, the impact of positive news or order executions remains uncertain, and available data for analysis is limited to historical prices and volumes, offering only a partial view of the market state. QT, as a sequential decision-making challenge, revolves around determining what and when to trade.

Within the realm of QT, the analytical structure evolves from a classic Markov Decision Process (MDP) to a Partially Observable Markov Decision Process (POMDP). An MDP is defined by a quintuple $\left[\mathcal{S}, \mathcal{A}, \mathcal{P}, \mathcal{R}, \gamma\right]$, where $\mathcal{S}$ denotes a distinct set of states, $\mathcal{A}$ represents a specific set of actions, $\mathcal{P}$ is the state transition probability function, and $\mathcal{R}$ signifies the reward function. By incorporating observational elements $O$ and $Z$, with $O$ being the observation set and $Z$ the observation probability function, the framework is modified into a POMDP. In this context, the agent receives various observations at each interval, and the comprehension of observable history until time $t$ is fundamental in capturing the dynamics of QT.

\textbf{Observation}. 
In the financial market, observations are split into two distinct categories: the portfolio observation subset $o_{p t}$ and the economic observation subset $o_{e t}$. The term $o_{p t}$ represents the aggregated profit of the portfolio, whereas $o_{e t}$ is associated with market pricing and technical metrics. The overall observation set $O$ comprises $O={P, I, R}$, leveraging technical indicators such as BuyLine and SellLine from the Dual Thrust strategy.

\textbf{Action}. Trading actions involve a continuous probability vector $a_t$ representing long and short positions. The agent executes the action associated with the maximum probability. Actions are represented as $a_t \in\left\{ \right. $ long $, $ short $\left.\right\}=\{1,-1\}$, simplifying position management and mitigating the impact of market capacity. Trading actions are interpreted as signals, guiding the execution of trades based on specific rules.

\textbf{Reward}: Practical constraints such as transaction fees and slippage are incorporated into the trading simulation. The account profit $r_t$ is computed considering these factors. To address the inefficiency of $r_t$ as a reward function, the Sharpe ratio is adopted. The Sharpe ratio ($Sr$) serves as a risk-adjusted return measure, representing the ratio of excess return to one unit of total risk.

This comprehensive framework allows for the representation of QT challenges in a dynamic and uncertain financial environment while incorporating realistic market constraints.

\section{observational and Recurrent DPG}

\begin{figure*}
    \centering
  \includegraphics[width=0.8\textwidth]{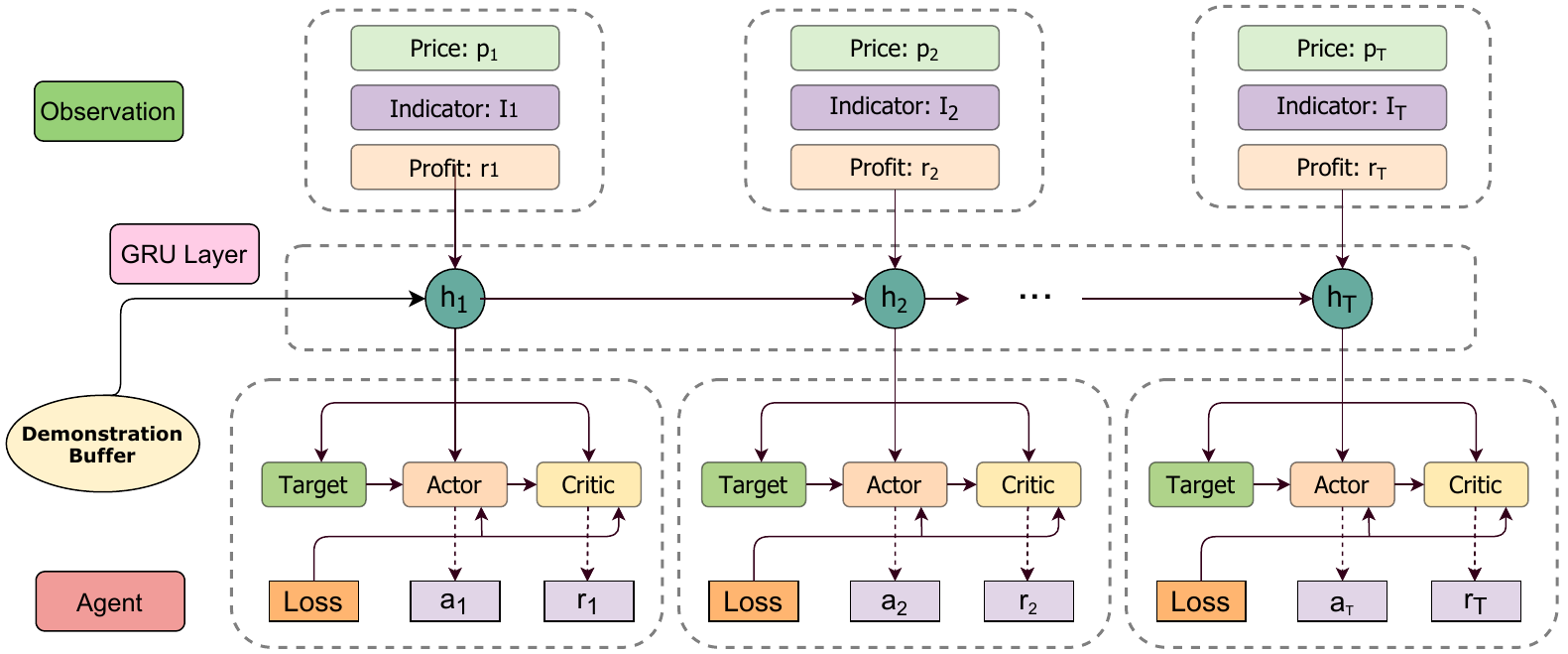}
  \caption{The overview of QTNet model}
  \label{flowchart_iRDPG}
\end{figure*}

In this segment, we introduce QTNet, our designed model, specifically devised for the POMDP setup in Quantitative Trading (QT). This model integrates elements from Recurrent Deterministic Policy Gradient (RDPG) and imitative learning in structured manner. Furthermore, a graphical representation of the QTNet architecture is provided in Figure \ref{flowchart_iRDPG}.

\subsection{Recurrent DPG}

Silver (2015) \cite{lillicrap2015continuous} unveiled the Deterministic Policy Gradient (DPG), a specialized off-policy reinforcement learning (RL) algorithm designed for continuous action spaces. Its application is especially pertinent to the domain of high-frequency Quantitative Trading (QT), where decisions need to be made on a continuum. The DPG's ability to handle the demands of constant trading activity, mindful of the costs associated with frequent transaction changes, makes it a suitable match for QT's requirements.
Building on DPG, the Recurrent Deterministic Policy Gradient (RDPG), introduced by Heess et al (2015) \cite{heess2015memory}, incorporates a recurrent structure to more adeptly navigate the QT landscape. This methodology acknowledges the importance of historical data sequences for QT, where the market's state isn't fully observable. Agents in this system utilize a historical compilation of market and personal account observations, including prices, indicators, and profits, encapsulated within a sequence $H$ denoted by $h_t=$ $\left(o_1, a_1, \ldots, o_{t-1}, a_{t-1}, o_t\right)$. RDPG employs recurrent neural networks (RNNs) to assimilate this history, enhancing the agent's ability to retain and utilize past information to inform future trades.

This approach aligns with the actor-critic methodology, where RDPG is adept at learning both a deterministic policy (actor) and an action-value function (critic). It aims to optimize the estimated action-value function $Q_\mu$, facilitated by the actor function $\mu(h)$ and critic function $Q(h, a)$, each parameterized by $\theta$ and $\omega$ respectively. A replay buffer is also maintained, archiving sequences of observations, actions, and rewards.

Our research innovates further by integrating Long Short-Term Memory (LSTM) networks into the QT framework, offering an alternative to Gate Recurrent Units (GRUs). LSTMs treat the prior observation-action history $h_{t-1}$ as the latent state from the previous timestep, leading to the current state $h_t$ being formulated as $h_t=$ $\operatorname{LSTM}\left(h_{t-1}, a_{t-1}, o_t\right)$. This substitution underscores the potential of LSTMs in capturing the temporal intricacies of the market, a crucial aspect for effective trading strategies within our POMDP model for QT.

\subsection{imitative learning based on DB and BC}

Effectively navigating the dynamic intricacies of financial market data poses a significant challenge due to the exponential growth of the exploration value space. Traditional model-free Reinforcement Learning (RL) algorithms face limitations in devising profitable policies efficiently within the context of Quantitative Trading (QT). Moreover, random exploration, lacking specific goals, becomes inefficient, especially when considering the imperatives of trading continuity and market friction factors. However, leveraging the model-free Recurrent Deterministic Policy Gradient (RDPG) algorithm with well-defined training goals proves to be a promising approach. As an off-policy algorithm, RDPG demonstrates adaptability to auxiliary data, providing the basis for the introduction of two pivotal components: the Demonstration Buffer and Behavior Cloning. These modules serve distinct roles in passive and active imitative learning, respectively,\cite{magdon2004maximum, murphy1999technical, zhao2017sequential,sutton2018reinforcement,ma2021global} strategically guiding our RDPG agent through the complex landscape of QT.

\textbf{Demonstration Buffer (DB)}. Initially, a prioritized replay buffer denoted as $DB$ is established, and it is pre-filled with demonstration episodes $\left(o_1, a_1, r_1, \ldots, o_T, a_T, r_T\right)$ obtained from the Dual Thrust strategy. Drawing inspiration from the methodologies introduced in DQfD Hester et al (2018) \cite{hester2018deep} and DDPGfD Vercerik (2017) \cite{vecerik2017leveraging}, our approach involves pretraining the agent using these demonstrations prior to engaging in the actual interaction. This pre-training, enriched with insights from technical analysis, equips the agent with a foundational trading strategy from the outset.

During the training phase, each minibatch consists of a mixture of instructional and agent-based episodes, selected using prioritized experience replay (PER) (Schaul et al. (2015) \cite{schaul2015prioritized}). PER promotes the selection of episodes with greater significance more frequently. The selection likelihood of an episode $P(i)$ is directly linked to its importance, defined as: $P(i)=\frac{p_t}{\sum_i p_i},$ where $p_i$ denotes the significance of episode $i$. In our implementation, we utilize the episode importance criteria as proposed by Vercerik (2017) \cite{vecerik2017leveraging}. Within this framework, $p_i$ is established as follows:

\begin{equation}
\mathbb{E}\left[\left|y_t^i-Q^\omega\left(h_t^i, a_t^i\right)\right|+\lambda_0\left|\nabla_a Q^\omega\left(h_t^i, a_t^i\right)\right|\right]+\epsilon_D,
\end{equation}

In this instance, the initial term signifies the episode's loss $L_i$, as detailed in Equation (8), and the subsequent term refers to the magnitude of the actor's gradient change, as shown in Equation (9). The constant $\theta_D$, designated for demonstration episodes, enhances their sampling likelihood, while $\lambda_0$ assesses the actor gradient's influence. With the alteration in sample allocation, network modifications are adjusted using significance sampling weights $w_i$:

\begin{equation}
w_i=\frac{1}{N} \times \frac{1}{P(i)^{\phi}}
\end{equation}
Here, $\phi$ replaces $\theta$ as the exponent, modifying the formulation slightly to ensure uniqueness.

Here, $\theta$ is a constant. This prioritized demonstration buffer controls the data ratio between demonstration and agent episodes. Importantly, it facilitates the efficient propagation of rewards.

\textbf{Behavior Cloning (BC)}. 
Behavior Cloning is utilized to define goals for every trading move. Intra-day opportunistic actions $(\bar{a})$  are integrated as expert-level maneuvers. Reflecting back, we develop a visionary trading expert, who invariably opts for a long position when prices are at their lowest and a short position at their peak. At each stage of training, the Behavior Cloning method (Ross and Bagnell (2010) \cite{ross2010efficient}) is applied to measure the differences between the agent's actions and the strategies of this foresighted expert.

Specifically, we employ Behavior Cloning losses ( $B C L o s s$ ) selectively when the critic $Q(h, a)$ indicates that the expert actions outperform the actor actions:
\begin{equation}
    L_{B C}=-E\left[\max \left(0, Q\left(h_t, \bar{a}_t\right)-Q\left(h_t, \mu_\theta\left(h_t\right)\right)\right)\right]
\end{equation}

This adjustment, referred to as Q-Filter (Nair et al. (2018) \cite{nair2018overcoming}), ensures that the Behavior Cloning losses are only considered when the expert actions are superior.

The Behavior Cloning loss $\left( L_{B C} \right )$ serves as an auxiliary loss for updates. Consequently, a modified policy gradient $\left(\nabla_\theta \bar{J}\right)$ is applied to the actor:

\begin{equation}
    \nabla_\theta \bar{J}=\lambda_1 \nabla_\theta J+\lambda_2 \nabla_\theta L_{B C}
\end{equation}

In this context, $\nabla_\theta J$ signifies the policy gradient as outlined in Equation (9), while $\lambda_1$ and $\lambda_2$ are parameters that govern the balance among the various loss functions. By integrating actions derived from expert insights, we set specific objectives for each stage of training, thereby minimizing phases of unproductive exploration. Behavior Cloning proves to be a valuable technique in aligning agent actions with expert strategies for improved trading performance.

\section{Experiments}

\begin{figure}[htbp]
    \centering
  \includegraphics[width=0.5\textwidth]{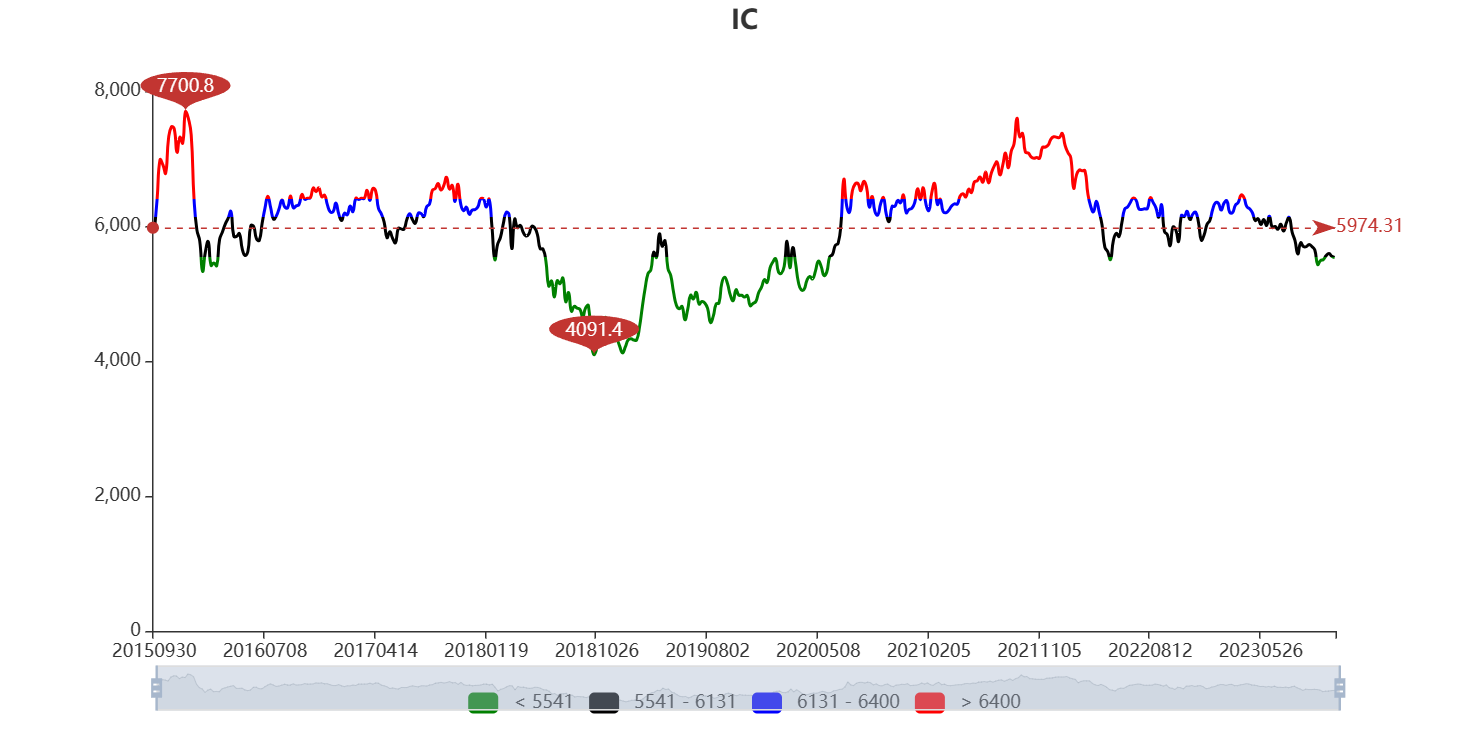}
  \caption{IC}
  \label{IC}
\end{figure}

\begin{figure}[htbp]
    \centering
  \includegraphics[width=0.5\textwidth]{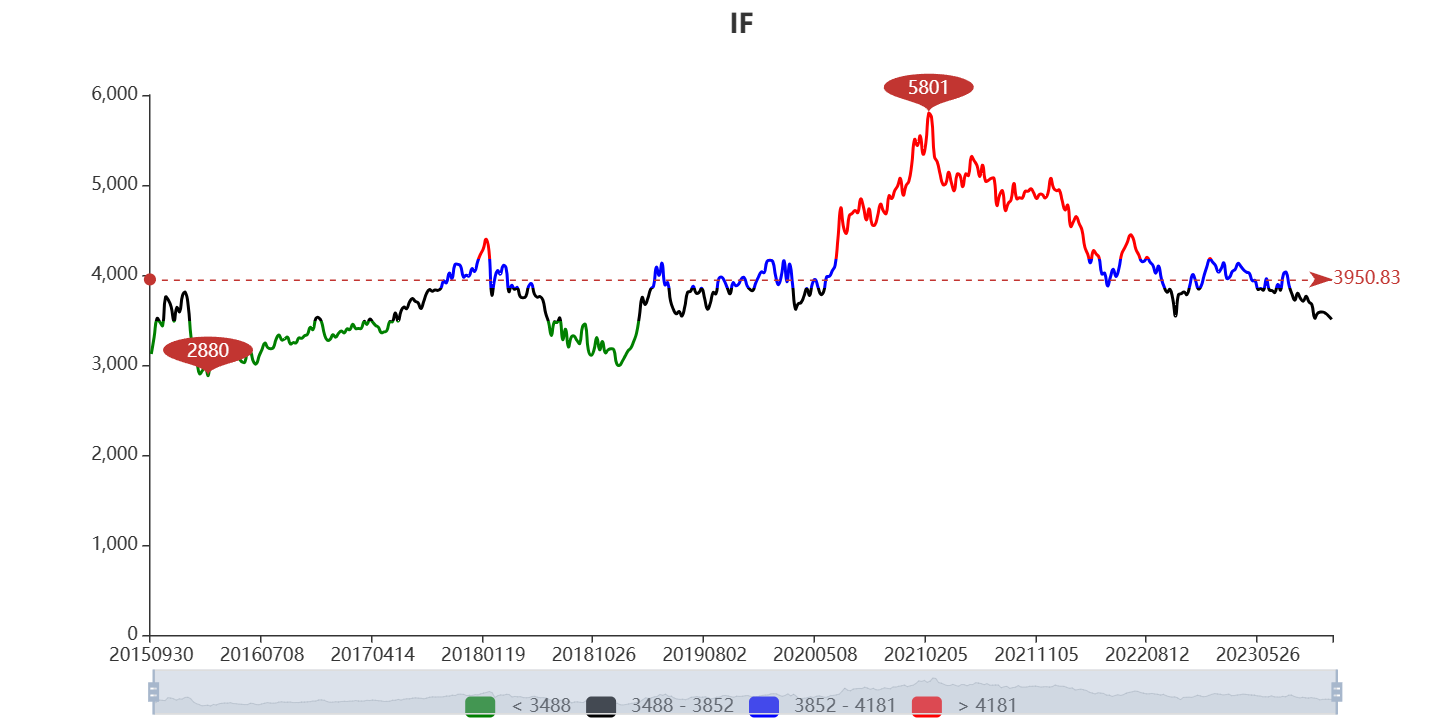}
  \caption{IF}
  \label{IF}
\end{figure}

In our empirical evaluation, we meticulously test the efficacy of our trading model by performing a back-test that incorporates data sampled at one-minute intervals from the Chinese financial futures market. This data encompasses two key stock-index futures: IF, which is constructed from a composite index reflecting the performance of the 300 foremost stocks traded on the Shanghai and Shenzhen stock exchanges, and IC, which similarly tracks an index but focuses on stocks with smaller market capitalizations. The back-testing process not only adheres to stringent, real-world trading conditions but also provides a detailed temporal resolution of market behavior. In illustration of our data's granularity, Figure \ref{IC} and Figure \ref{IF} present a time series of the closing prices recorded on a minute-by-minute basis for both the IF and IC futures, thereby offering a precise visual account of their market movements during the testing period.

\subsection{Setup}
In our study, we deploy a data-driven approach, utilizing minute-bar OHLC (Open, High, Low, Close) data for futures trading, which captures the nuances of price movements within each minute. This high-frequency data, commonly found in actual financial markets, poses unique challenges for reinforcement learning $(R L)$ algorithms, particularly in maintaining action continuity. The dataset, sourced from python package $'qstock'$, which is a quantitative investment analysis open source library, currently includes data acquisition, visualization, stock selection and quantitative backtest four modules. Among them, the data module comes from the open data of the Oriental Wealth network, Flush, Sina Finance, and other online resources. 

Data spans a training period from September 28, 2015, to November 10, 2022, and a subsequent testing period from November 11, 2022, to November 10, 2023. To mirror real-market conditions more accurately, we integrate practical elements like a transaction fee (notated as $\delta=2 \times 10^{-5}$ ) and a fixed slippage amount $(\zeta=0.15)$.

For the purposes of our simulation, certain assumptions are made. We presume that each order is executed right at the opening of each minute bar, with the rewards calculated at the minute's closing. Key futures-specific considerations, such as margin requirements and contract settlement peculiarities, are factored into our model. The training process is designed to conclude under two conditions: either a $50 \%$ loss of positions or the depletion of the required margin. We set the initial account balance at $\$ 1,000,000$ for the commencement of the testing phase.

The performance of our trading policy is assessed using a suite of standard quantitative trading (QT) metrics:

\begin{itemize}
  \item Total Return Rate (Tr): Calculated as $\left(P_{\text {end}} - P_{\text {start}}\right) / P_{\text {start}}$ where $P$ represents the total value of the position and cash.
  
  \item Sharpe Ratio (Sr): Originally introduced by Sharpe (1966), it is defined as $E[r] / \sigma[r]$, providing a measure of excess return over unit systematic risk.
  
  \item Volatility (Vol): Expressed by $\sigma[r]$, with $r$ representing the historical series of return rates, this metric illustrates the variability of returns and the associated risk factors of the strategies.
  
  \item Maximum Drawdown (Mdd): Determined by $\max \left(P_i-P_j\right) / P_i$ for $j>i$, this metric quantifies the most significant historical reduction, signifying the most adverse scenario encountered.
  
\end{itemize}

These metrics collectively offer a comprehensive evaluation of our strategy, with $\mathrm{Tr}$ providing a direct gauge of profitability, Sr and Vol offering insights into risk-adjusted returns and the stability of returns, and Mdd presenting a measure of potential worst-case scenarios.

\subsection{imitative learning Strategies}
In our research, we incorporate the Dual Thrust strategy into the Demonstration Buffer module as a model trading policy. This strategy, rooted in the principles of technical analysis, particularly oscillators, sets a rational price oscillation range using a calculation based on the highest high (HH), lowest close (LC), highest close (HC), and lowest low (LL) of the past n periods. The formula for the range is defined as the maximum of either HH minus LC or HC minus LL. On each trading day, two key thresholds, the BuyLine, and the SellLine, are established by respectively adding and subtracting a specified proportion of this Range to the day's opening price, where K1 and K2 are constants dictating the market's resistance levels for breaching these thresholds.

The Dual Thrust strategy creates trading signals when the current price breaches the Range's threshold, either upwards or downwards. In the context of our Behavioral Cloning module, we also introduce intra-day greedy actions as the benchmark for expert behavior. This involves taking long positions at the day’s lowest price and short positions at the highest, mimicking a theoretically optimal, albeit opportunistic, strategy. These actions provide a vital reference for training the agent, with the expert action at each time step determined by the minimum or maximum opening prices over a trading day.

For every trading session, the upper boundary, termed BuyLine, and the lower boundary, known as SellLine, are calculated. This is done by either adding or subtracting a specific percentage of the Range to or from the opening price of the day. The key components of the Dual Thrust strategy are expressed as follows:

\begin{equation}
\begin{aligned}
& \text { Range }=\max [H H-L C, H C-L L], \\
& \text { BuyLine }=\text { Open }+K_1 \times \text { Range, } \\
& \text { SellLine }=\text { Open }-K_2 \times \text { Range, }
\end{aligned}
\end{equation}

Here, Open represents the day's opening price, and $K_1$ and $K_2$ are constants controlling the resistance levels of market prices against breaking BuyLine and SellLine, respectively. Additionally, denotes the highest high price over the previous $n$ time periods.

In practical terms, trading signals are generated based on the Dual Thrust strategy when the current price exceeds a certain percentage of the Range either upwards or downwards.

Moving into the Behavioral Cloning module, intra-day greedy actions are introduced as expert actions from a hindsight perspective. Adopting a long position at the lowest price and a short position at the highest price consistently represents a relatively optimal and greedy strategy. This prophetic policy serves as expert actions for the agent throughout the training process. At each time step $t$, the expert action is determined by:
$$
\bar{a}_t= \begin{cases}1, & \text { if } t=\arg \min P_{o_{t-n d t}} \\ -1, & \text { if } t=\arg \max P_{o_{t-n d t}}\end{cases}
$$

Here, $n d$ denotes the length of one trading day, and $P_{o_{t-n d t}}$ represents the sequence of opening prices for the day. This approach provides a rich foundation for understanding and incorporating expert behavior into the agent's learning process.

Our study also includes a comparison of the proposed Iterative Recurrent Deep Policy Gradient (QTNet) method against several baseline strategies for context. These include the Long \& Hold strategy, where a long position is maintained throughout the test period, and its counterpart, the Short \& Hold strategy. Additionally, we examine the Deep Deterministic Policy Gradient (DDPG) algorithm, known for its effectiveness in continuous control scenarios, as a comparative benchmark. This juxtaposition allows us to evaluate the effectiveness of our approach in the dynamic and often unpredictable realm of quantitative trading.

\subsection{Baseline Methods}
Our proposed QTNet is subjected to comparison with several baseline strategies:

\begin{itemize}
    \item  Long \& Hold: This strategy involves initiating a long position at the outset and maintaining the position until the conclusion of the test period. 
    \item Short \& Hold: In contrast to the Long \& Hold strategy, Short \& Hold commences with taking a short position at the beginning and retaining the position throughout the test period.
    \item DDPG (Lillicrap et al. (2015) \cite{lillicrap2015continuous}): DDPG stands for Deep Deterministic Policy Gradient, an off-policy model-free actor-critic reinforcement learning algorithm. Typically, it exhibits robust performance in tasks characterized by continuous control.
\end{itemize}

\subsection{Experimental Outcomes}
In this section, we present a comprehensive set of experiments to compare the performance of our QTNet model against baseline methods. Ablation experiments are also conducted to highlight the significance of each module within QTNet. To evaluate the generalization capabilities of both QTNet and the Dual Thrust strategy, tests are performed across distinct futures markets.

\subsection{Data Demonstration}
In our study, we utilized minute-level frequency data for IC futures to assess the performance of the QTNet model against several baseline trading strategies. As detailed in Table \ref{performance_comparison}, the QTNet consistently outperforms traditional methods, particularly in terms of the total return rate (Tr) and Sharpe Ratio (Sr), indicating its adaptability and profitability in a high-frequency quantitative trading setting. Although DDPG is a renowned reinforcement learning algorithm that excels in certain scenarios, it displayed relatively lower performance in our tests, which may be attributed to its adaptability in handling high-frequency trading data.

The QTNet also demonstrated impressive performance in the critical risk management metric of Maximum Drawdown (Mdd), showing more robustness during market downturns compared to straightforward strategies such as Long \& Hold and Short \& Hold, thereby minimizing potential maximum losses. This result highlights the resilience of the QTNet strategy in the face of market volatility and its capacity to maintain stability in complex market conditions. Overall, the comprehensive performance of QTNet validates the efficiency of recurrent GRU networks in capturing the complexities of market dynamics in quantitative trading.

\begin{table}[htbp]
\centering
\caption{Performance of comparison methods on IC.}
\label{performance_comparison}
\begin{tabular}{@{}lcccc@{}}
\toprule
\textbf{Methods} & \textbf{Tr (\%)} & \textbf{Sr} & \textbf{Vol} & \textbf{Mdd (\%)} \\ \midrule
Long \& Hold    & -8.32          & -0.318       & 0.746        & 62.84             \\
Short \& Hold   & 9.53            & 0.167        & 0.658        & 49.30             \\
DDPG            & -17.26          & -0.428       & 0.498       & 57.22             \\
QTNet           & 20.28           & 0.562        & 0.421        & 23.73             \\ \bottomrule
\end{tabular}
\end{table}

\subsection{Ablation Experiments}
The ablation studies focusing on IC futures shed light on the significance of each element within the QTNet framework concerning its aggregate performance. The data extracted from Table \ref{tbl:ablation} demonstrates that the integrated return trajectory of IR-DPG stands out against its counterparts for the duration of the evaluation phase. This study dissects the IR-RDPG into three distinct iterations: the basic RDPG utilizing only GRU networks, RDPG-DB which integrates a demonstration buffer, and RDPG-BC which employs a behavior cloning module. The analysis delineates that the base RDPG variant grapples with achieving action coherence and mastering a lucrative trading approach. The assimilation of the demonstration buffer in RDPG-DB markedly augments the efficiency of experience sampling, while the behavior cloning feature in RDPG-BC curtails the detrimental effects of random exploration on the agent's performance. The fully-fledged QTNet, amalgamating all the components, evidently excels in terms of both profit generation and risk management metrics, underlining the prowess of imitative learning strategies in the domain.

\begin{table}[htbp]
\centering
\caption{Ablation experiments on IC.}
\label{tbl:ablation}
\begin{tabular}{lcccc}
\toprule
Methods   & Tr (\%) & Sr    & Vol   & Mdd (\%) \\
\midrule
RDPG      & 8.96    & 0.067 & 0.590 & 34.62    \\
RDPG-DB   & 18.72   & 0.467 & 0.452 & 23.72    \\
RDPG-BC   & 31.24   & 0.619 & 0.427 & 29.81    \\
QTNet     & 36.26   & 0.742 & 0.411 & 25.37    \\
\bottomrule
\end{tabular}
\end{table}

\subsection{Generalization Ability}
The versatility of the QTNet model is assessed by its performance in varying market conditions, specifically in IF and IC futures, as depicted in Table \ref{tbl:performance_comparison}. The table elucidates the stark contrast in Dual Thrust's effectiveness between the IC and IF markets, which illustrates the potential constraints of fixed trading indicators. On the other hand, QTNet exhibits superior performance, especially in the IC market, regardless of being trained solely on IF data. This demonstrates QTNet's inherent flexibility and its proficiency in discerning durable features that are crucial for formulating dynamic trading strategies applicable to different market environments. The model's ability to perform well across various markets cements its standing as a robust and adaptable tool for financial applications in the real world.

\begin{table}[htbp]
\centering
\caption{Performance of comparison methods on IC and IF.}
\label{tbl:performance_comparison}
\begin{tabular}{lccccc}
\toprule
Methods     & Data & Tr(\%) & Sr    & Vol   & Mdd(\%) \\
\midrule
Dual Thrust & IC   & 26.40  & 0.810 & 0.469 & 19.24   \\
            & IF   & -29.59 & -0.577& 0.796 & 55.81   \\
QTNet       & IC   & 34.27  & 0.742 & 0.457 & 24.56   \\
            & IF   & 28.73  & 0.523 & 0.517 & 26.25   \\
\bottomrule
\end{tabular}
\end{table}

\section{Conclusion}
In our research, we unveiled the QTNet, a dynamic, imitative learning framework designed for the challenges of Quantitative Trading (QT). We constructed a Partially Observable Markov Decision Process (POMDP) to adeptly handle the erratic nature of high-frequency trading data. Crucially, our model harnesses the principles of imitative learning to harmonize the exploration-exploitation trade-off, thus refining the decision-making process of our trading agent. We validated the performance of QTNet by conducting rigorous evaluations using authentic data from stock-index futures, acknowledging the operational limitations typical of financial trading. The outcomes affirmed the financial viability and risk mitigation prowess of QTNet. Through a series of comparative analyses, the model's capacity to adapt to varied market conditions was put on display. To conclude, QTNet underscores the advantage for trading agents in real-world markets to draw on established trading techniques. Our results underscore the enhanced adaptability and performance gains achieved by merging imitative learning with a reinforcement learning paradigm in the realm of QT solutions.

\bibliographystyle{plain}
\bibliography{reference}

\end{document}